\documentclass[12pt]{article}
\usepackage{dsfont}
\usepackage{amssymb}
\usepackage{graphicx}

\begin{document}
\title{Bogoliubov theory of the Hawking effect in Bose-Einstein condensates}
\author{
U. Leonhardt$^{1}$, T. Kiss$^{1,2,3}$, and P. \"Ohberg$^{1,4}$\\
$^{1}$School of Physics and Astronomy, University of St Andrews,\\
North Haugh, St Andrews KY16 9SS, Scotland\\
$^{2}$Research Institute for Solid State Physics and Optics,\\ 
H-1525 Budapest, P.~O.~Box 49, Hungary\\ 
$^{3}$Institute of Physics, University of P\'ecs,\\
Ifj\'us\'ag u.\ 6. H-7624 P\'ecs, Hungary\\
$^{4}$Department of Physics, University of Strathclyde,\\
Glasgow G4 0NG, Scotland}
\date{}
\maketitle
\begin{abstract}
Artificial black holes may demonstrate some of the elusive quantum 
properties of the event horizon, in particular Hawking radiation.
One promising candidate is a sonic hole in a Bose-Einstein condensate.
We clarify why Hawking radiation emerges from the condensate and 
how this condensed-matter analog reflects some of the intriguing 
aspects of quantum black holes.
\end{abstract}

\newpage

\section{Introduction}

Picture a Bose-Einstein condensate flowing through a nozzle where the
condensate exceeds the speed of sound. Suppose that the nozzle is designed
such that the transsonic flow does not become turbulent. One could build such
a nozzle, the equivalent of the Laval nozzle \cite{Courant}, out of light, using
the dipole force between light and atoms to confine the condensate in an
appropriate potential. Consider the fate of sound waves propagating against
the current of the transsonic condensate. In the subsonic region sound waves
may advance against the flow, whereas in the supersonic zone they are simply
swept away. No sound can escape the point where the flow turns supersonic
-- the sonic horizon. The transsonic fluid acts as the acoustic equivalent of
the black hole \cite{Unruh,Visser}.

An artificial black hole \cite{Book} of this kind could be employed to
demonstrate some elusive quantum properties of the event horizon in the
laboratory, in particular Hawking radiation \cite{BD,Brout}. Hawking
\cite{Hawking} predicted that the event horizon emits quanta as if the horizon
had a temperature given by the gradient of the gravitational potential. To be
more precise, the horizon should spontaneously emit quantum pairs where one
particle of each pair falls into the hole and the other escapes into
space, constituting the radiation of the horizon. Both the spectral
distribution and the quantum state of the emerging radiation are thermal.
For solar-mass or larger black holes the Hawking temperature is in the order
of $10^{-7}$ K or below, which makes the effect next to impossible to observe
in astronomy. In the case of sonic holes the Hawking temperature is given 
by the velocity gradient $\alpha$ at the sonic horizon \cite{Unruh,Visser},
%%%%%%%
\begin{equation}
  \label{eq:hawking}
  k_{_B} T = \frac{\hbar \alpha}{2 \pi} \, ,
\end{equation}
%%%%%%%
and the emitted quanta are phonons. A velocity gradient of $10^3$ Hz would
correspond to about $1.2$ nK temperature. In order to observe such subtle
quantum effects one should employ the best and coldest superfluids available ---
Bose-Einstein condensates of dilute gases \cite{Dalfovo}.

Some detailed schemes for sonic black holes in Bose-Einstein condensates have
been investigated theoretically 
\cite{Garay1,Garay2,Barcelo1,Barcelo2,Sakagami}. 
The ultimate design depends on
experimental details and on the state of the art in manipulating condensates,
a rapidly evolving field. In this paper we analyze the general aspects of the
Hawking effect in Bose-Einstein condensates. 
In the first part of the paper we collect and combine the ingredients
of the effect, results that are scattered in the literature.
In the second part we show how the Hawking effect arises naturally 
within the Bogoliubov theory of the elementary excitations in
Bose-Einstein condensates \cite{Fetter1,Fetter2}.
For the first time, to our knowledge, we connect the quantum physics 
of the event horizon to the behavior of a realistic quantum fluid.

\newpage
\section{Sound in fluids}

Consider the propagation of sound in fluids moving with flow speed $\bf
u$. Suppose that the flow varies little over the scale of an acoustic
wavelength. In this regime we can describe sound propagation in geometrical
acoustics (the equivalent of geometrical optics \cite{BornWolf} or of the
semiclassical approximation in quantum mechanics \cite{LL3}). 
Sound rays follow Hamilton's equations,
%%%%%%%
\begin{equation}
  \frac{d {\bf r}}{d t} = \frac{\partial \omega}{\partial {\bf k}} \, , \quad
  \frac{d {\bf k}}{d t} = - \frac{\partial \omega}{\partial {\bf r}} \, ,
\end{equation}
%%%%%%%
where the dispersion relation between the frequency $\omega$ and the wave
vector $\bf k$ defines the effective Hamiltonian $\omega ({\bf r}, {\bf k})$.
Assume that in each fluid cell
%%%%%%%
\begin{equation}
  \label{eq:disrel}
  \omega^{\prime 2} = c^2 k^2 \, ,
\end{equation}
%%%%%%%
where $c$ denotes the speed of sound and $\omega'$ refers to the frequency in
locally comoving frames. In the laboratory frame, $\omega'$ is
Doppler-shifted,
%%%%%%%
\begin{equation}
  \label{eq:doppler}
  \omega' = \omega - {\bf u} \cdot {\bf k} \, .
\end{equation}
%%%%%%%
In order to see why waves in fluids are related to waves in general
relativity, we write the dispersion relation (\ref{eq:disrel}) in a relativistic
form. We introduce the space-time wave vector
%%%%%%%
\begin{equation}
  k_\nu = (- \omega, {\bf k})
\end{equation}
%%%%%%%
and the matrix 
%%%%%%%
\begin{equation}
  g^{\mu \nu} = \Omega^{-2}\left( 
    \begin{array}{cc}
      1 & {\bf u} \\
      {\bf u} & -c^2 \mathds{1} + {\bf u} \otimes {\bf u}
    \end{array}
\right) \, .
\end{equation}
%%%%%%%
The prefactor $\Omega$ is an arbitrary non-negative function of the
coordinates called the conformal factor. 
In this notation the dispersion relation appears in the relativistic form
%%%%%%%
\begin{equation}
  g^{\mu \nu} k_\mu k_\nu = 0 \, ,
\end{equation}
%%%%%%%
adopting Einstein's summation convention. Therefore, sound waves experience
the moving fluid as an effective space-time geometry with the metric $g_{\mu
  \nu}$, the inverse matrix of $g^{\mu \nu}$, given by
%%%%%%%
\begin{equation}
  g_{\mu \nu} = \Omega^2 
  \left(
    \begin{array}{cc}
      c^2 - u^2 & {\bf u} \\
      {\bf u} & - \mathds{1}
    \end{array}
  \right) \, .
\end{equation}
%%%%%%%
The analogy between sound waves in fluids and waves in general relativity
turns out to be exact for an irrotational fluid \cite{Remark1} with arbitrary
density profile $\rho_0$, flow ${\bf u}$ and speed of sound $c$, where
$\rho_0$, ${\bf u}$ and $c$ may vary in space and time.
The velocity potential $\varphi$ and the density perturbations $\rho_s$ of
sound obey the linearized equation of continuity and the linearized
Bernoulli equation \cite{LL6}
%%%%%%%
\begin{eqnarray} 
 \partial_t \rho_s + \nabla \cdot (\mathbf{u} \rho_s + \rho_0 \nabla
 \varphi ) &=& 0\,, \label{eq:cont} \\
 (\partial_t + \mathbf{u} \cdot \nabla ) \varphi + c^2
 \frac{\rho_s}{\rho_0} &=& 0 \,. \label{eq:bern}
\end{eqnarray}
%%%%%%%
As a consequence, the velocity potential $\varphi$ of sound
obeys the equation \cite{Visser}
%%%%%%%
\begin{equation}
 \label{eq:wave}
  \partial_t \frac{\rho_0}{c^2} (\partial_t + \mathbf{u} \cdot \nabla ) \varphi
  + \nabla\cdot  \frac{\rho_0}{c^2}
     \left[{\bf u}\,\partial_t -(c^2-u^2)\right]\varphi = 0 \,,
\end{equation}
%%%%%%%
which can be written as the relativistic wave equation \cite{Unruh,Visser}
%%%%%%%
\begin{equation}
  \label{eq:waverel}
  D_\nu D^\nu \varphi = \frac{1}{\sqrt{-g}} \partial_\mu \sqrt{-g} g^{\mu \nu}
  \partial_\nu \varphi = 0
\end{equation}
%%%%%%%
with the conformal factor $\Omega$, in $d$ spatial dimensions, 
chosen as \cite{Visser}
%%%%%%%
\begin{equation}
  \Omega = \left( \frac{\rho_0}{c^3} \right)^{1-d} \, .
\end{equation}
%%%%%%%
The assumptions made in order to derive the wave equation (\ref{eq:wave}) are:
The fluid \cite{LL6} is irrotational (1) and isentropic (2).
Bose-Einstein condensates naturally satisfy condition (1).
Condition (2) characterizes the hydrodynamic regime of
condensates \cite{Dalfovo}. 
Here the local pressure depends only on the density and on the temperature 
of the fluid and the quantum pressure is negligible \cite{Dalfovo}. 
Condition (2) turns out to be violated close to the sonic horizon. 

\section{Sonic horizon}

Consider the propagation of sound waves in the vicinity of the sonic
horizon. Focus on the physics in the direction $z$ of the flow at the horizon
in a quasi-onedimensional model. Assume that the speed of sound in the fluid
is constant. The wave equation (\ref{eq:wave}) reads explicitly
%%%%%%%
\begin{equation}
  \partial_t (\partial_t + u \partial_z) \varphi + \partial_z \left[ 
    u \partial_t - (c^2- u^2) \partial_z \right] \varphi = 0 \, .
\end{equation}
%%%%%%%
We obtain the general solution
%%%%%%%
\begin{equation}
  \label{eq:s}
  \varphi= \varphi_0 (\tau_{\pm} -t) \, , \quad
  \tau_{\pm} = \int \frac{dz}{c \pm u} \, .
\end{equation}
%%%%%%%
The $\tau_{\pm} - t$ refer to null coordinates in the frame comoving with the
fluid \cite{Remark2}. In these coordinates sound waves propagate exactly like
in homogenous space. In the laboratory frame sound waves are accelerated or
slowed down by their carrier, the moving fluid. An interesting behavior occurs
near the horizon, say at $z=0$, where
%%%%%%%
\begin{equation}
  \label{eq:alpha}
  u = - c + \alpha z \, .
\end{equation}
%%%%%%%
The constant $\alpha$ describes the velocity gradient of the condensate at the
sonic horizon. We see that
%%%%%%%
\begin{equation}
  \label{eq:tauplus}
  \tau_+ = \frac{\ln (z/z_\infty)}{\alpha} \, .
\end{equation}
%%%%%%%
Wave packets localized just before the horizon at $z \gtrsim 0$
take an exponentially large time to advance against the current. On the other
side of the horizon, $z \lesssim 0$, such waves  drift equally slowly in the
direction of the flow. The horizon at $z=0$ marks a clear watershed, cutting
space into two disconnected regions. In terms of the coordinate 
(\ref{eq:tauplus}) these regions are characterized by the sign of $z_\infty$.
For stationary sound waves with frequency $\omega$ we get from the 
general solution (\ref{eq:s}) 
%%%%%%%
\begin{equation}
  \label{eq:stat}
  \varphi= \mbox{Re} \left\{ 
     \varphi_{_A} z^{i \omega/\alpha} e^{-i \omega t} \right\} \, .
\end{equation}
%%%%%%%
The phase of the wave, $(\omega/\alpha) \ln (z/z_\infty)$, diverges
logarithmically at the horizon where, in turn, the wavenumber $k$ develops a
pole,
%%%%%%%
\begin{equation}
  \label{eq:kz}
  k = \frac{\partial}{\partial z} \left( \frac{\omega}{\alpha} \ln
  \frac{z}{z_\infty} \right) = \frac{\omega}{\alpha z} \, ,
\end{equation}
%%%%%%%
and the wavelength of sound shrinks beyond all scales,
%%%%%%%
\begin{equation}
  \lambda = \frac{2 \pi}{ k} = \lambda_0 \alpha z \, .
\end{equation}
%%%%%%%
However, when $\lambda$ reaches the scale of the healing length $\xi$ of the
condensate \cite{Dalfovo} (also called the correlation length) the
hydrodynamic description of sound in Bose-Einstein condensates is no longer
valid \cite{Dalfovo}. The acoustic theory at the horizon predicts its own
demise. Similarly, waves near the event horizon of a gravitational black hole
are compressed beyond all scales. New physics beyond the Planck
scale may affect the event horizon \cite{T,J,U,BMRS}.

\section{Bogoliubov dispersion}

For Bose-Einstein condensates the equivalent of trans-Planckian physics is
well-known --- Bogoliubov's theory of elementary excitations. In the dispersion
relation (\ref{eq:disrel}) we replace the right-hand side by Bogoliubov's
famous result \cite{Dalfovo,Bogo,LL9,CJ0}
%%%%%%%
\begin{equation}
  \label{eq:bogo}
  \omega^{\prime 2} = c^2 k^2 \left( 1 + \frac{k^2}{k_c^2}\right) \, .
\end{equation}
%%%%%%%
The parameter $k_c$ is the acoustic Compton wave number
%%%%%%%
\begin{equation}
  k_c = \frac{m c}{\hbar} = \frac{1}{\xi \sqrt{2}} \, ,
\end{equation}
%%%%%%%
with $m$ being the atomic mass, expressed also 
in terms of the healing length $\xi$ \cite{Dalfovo} 
(the correlation length). 
Typically, $\xi$ is in the order of $10^{-6}\,{\rm m}$ and 
$c$ reaches a few $10^{-3}\,{\rm m}/{\rm s}$ 
in Bose-Einstein condensates
(without exploiting Feshbach resonances).
We calculate the group velocity
%%%%%%%
\begin{eqnarray}
  \label{eq:group}
  v&=&\frac{\partial \omega}{\partial k} = u+v' \, , \\
  \label{eq:groupprime}
  v'&=&\frac{\partial\omega'}{\partial k} = c^2\frac{k}{\omega'}
   \left(1+\frac{2k^2}{k_c^2}\right)\, .
\end{eqnarray}
%%%%%%%
Equation (\ref{eq:group}) shows that the group velocity obeys 
the Galilean addition theorem of velocities.
Equation (\ref{eq:groupprime}) expresses the group velocity 
in the fluid frame, $v'$, in terms of the frequency and the wavenumber.
The acoustic Compton wavenumber, $k_c$, sets the scale beyond 
which $v'$ deviates significantly from $c$. For large wavelengths, 
sound is communicated by atomic collision, and the product of the 
condensate's density and the atomic collision strength gives 
$mc^2$ \cite{Dalfovo}.
For wavelengths comparable with or shorter than the healing length,
the interaction-free Schr\"odinger dynamics of the atoms dominates
the transport of excitations. Perturbations of the free wavefunction 
travel with infinite velocity. So the acoustic Compton wavenumber, 
$k_c$, characterizes the crossover between
the speed of sound and the infinite speed of perturbations 
of free matter waves.

Close to the sonic horizon, the wavenumber (\ref{eq:kz}) 
increases dramatically, and, in turn, the effective speed of sound $v'$ grows. 
The horizon, defined as the place where the fluid exceeds the speed
of sound, seems to dissolve like a mirage.
Nature appears to prevent the existence of an event horizon.
However, we show in Sec.\ 6 that the horizon still exists, 
but at a less well-defined location and 
for a particular class of elementary excitations only.
As long as $k^2$ is much smaller than $k_c^2$ we get the acoustic relation
%%%%%%%
\begin{equation}
  \label{eq:acoustic}
  k \sim \frac{\omega}{u\pm c} \, .
\end{equation}
%%%%%%%
For the other extreme, where $k^2$ is much larger than $k_c^2$, 
one finds \cite{Corley,CJ}
%%%%%%%
\begin{equation}
  \label{eq:trans}
  k \sim \pm 2 k_c \sqrt{u^2/c^2-1}+ \frac{\omega u}{ c^2 - u^2} \, .
\end{equation}
Consider the turning points $z_0$, the points where
the group velocity of sound (\ref{eq:group}) vanishes.
If the acoustic dispersion relation (\ref{eq:disrel}) were universally valid
the horizon would be the turning point. Therefore $|z| \lesssim |z_0|$ does
indicate the spatial scale of the trans-acoustical range around the horizon,
which defines the spatial delocalization of the horizon.
To proceed we recall that elementary excitations are small perturbations of
the condensate. Their energies $\hbar \omega$ ought to be much smaller than 
the mean-field energy of the condensate, 
which is in the order of $mc^2$ (with $c$ being the speed of sound).
Therefore, 
%%%%%%%
\begin{equation}
  \label{eq:eps}
  \varepsilon= \frac{\hbar \omega}{m c^2}\,,\quad |\varepsilon|\ll 1\,.
\end{equation}
%%%%%%%
We expand the solution $z_0$ of $v=0$ as a power series
in $\varepsilon^{1/3}$ and find, to leading order, three turning points in the
complex plane given by \cite{LKO}
%%%%%%%
\begin{equation}
  \label{eq:z0}
  z_0 = \frac{c}{\alpha} \frac{3}{2} \sqrt[3]{-1} \left( \frac{\varepsilon}{2}
  \right)^{2/3} \, .
\end{equation}
%%%%%%%
Far away from the horizon we may characterize the four fundamental solutions
of the dispersion relation (\ref{eq:bogo}) combined with the Doppler shift
(\ref{eq:doppler}) by their asymptotics (\ref{eq:acoustic}) or
(\ref{eq:trans}).  However, close to a turning point geometrical
acoustics alone does not provide a good description of wave propagation
anymore. The turning points may cause scattering.  The connections between
elementary excitations across the horizon must be examined with care.
Extending $z$ to the complex plane represents an elegant way of analyzing this
connection. We find in the Appendix that the acoustic relation (\ref{eq:kz})
remains valid on either the upper or the lower half of the complex plane.

\section{Bogoliubov modes}

Elementary excitations are perturbations of the condensate, ripples on the
macroscopic wave function $\psi_0$ of the condensed atoms. The excitations
constitute the non-condensed part of the atomic gas. To describe elementary
excitations, the total many-body field operator $\hat \psi$ of the atoms is
split into two components, the condensate with the mean-field wave function 
$\psi_0$ and the non-condensed part. 
The mean-field wave function comprises the density profile $\rho_0$
and the flow ${\bf u}$, as
%%%%%%%
\begin{equation}
  \psi_0=\sqrt{\rho_0}\, e^{iS_0} \, ,\quad 
  {\bf u} = \frac{\hbar}{m}\nabla S_0 \,.
\end{equation}
%%%%%%%
The atomic field operator $\hat \psi$ is split into the condensate and
the non-condensed part according to the relation
%%%%%%%
\begin{equation}
  \hat \psi = \psi_0 + e^{i S_0} \hat \phi \, .
\end{equation}
%%%%%%% 
The non-condensed part consists of Bogoliubov modes $u_\nu$ and $v_\nu$
\cite{Dalfovo,Fetter1,Fetter2},
%%%%%%%
\begin{equation}
  \hat \phi = \sum_\nu \left( u_\nu \hat a_\nu+ v_\nu^* \hat a _\nu^\dagger
  \right) \, .
\end{equation}
%%%%%%%
The $u_\nu$ and $v_\nu$ are subject to the Bogoliubov-deGennes equations
\cite{Dalfovo,Fetter1,Fetter2}. If one requires that the Bogoliubov modes satisfy 
the orthonormality relations
%%%%%%%
\begin{eqnarray}  
  \int (u_\nu^* u_{\nu'} - v_\nu^* v_{\nu'} ) \, dz &=& \delta_{\nu \nu'} \, ,
  \label{eq:norm}  \\
  \int (u_\nu v_{\nu'} - v_\nu u_{\nu'}) \, dz &=& 0 \, , \label{eq:zero} 
\end{eqnarray}
%%%%%%%
then the $\hat a_\nu$ and $\hat a^\dagger_\nu$ obey the commutation relations
of Bose annihilation and creation operators, as a consequence of  the 
fundamental commutator of atoms with Bose statistics
%%%%%%% 
\begin{equation}
  \left[ \hat \psi (z) , \hat \psi^\dagger (z') \right] = \delta (z-z') \, .
\end{equation}
%%%%%%% 
Each pair of $u_\nu$ and $v_\nu$ characterizes the spatial shape and the
evolution of an excitation wave, and the Fock space of the $\hat a_\nu$
and $\hat a_\nu^\dagger$ spans the state space of the excitation
quasiparticles.

To see how the Bogoliubov modes are related to sound waves, we write down the
macroscopic wave function of the condensate combined with 
one of the excitations,
%%%%%%%
\begin{equation}
  \psi = \psi_0 + e^{i S_0} (u_\nu + v_\nu^* ) \, .
\end{equation}
%%%%%%%
We represent $\psi$ as
%%%%%%%
\begin{equation}
  \psi = \sqrt{\rho}\, e^{iS} \, , \quad
  \rho = \rho_0 + \rho_s \, , \quad
   S= S_0 +s \, ,
\end{equation}
%%%%%%%
where $\rho_s$ denotes the local density of the sound wave and $s$ is
proportional to the velocity potential
%%%%%%%
\begin{equation}
  \varphi = \frac{\hbar}{m}s  \, .
\end{equation}
%%%%%%%
Assuming that $\rho_s$ and $s$ are
small perturbations we get
%%%%%%%
\begin{equation}
  u_\nu +v_\nu^* = \sqrt{\rho_0} \left( \frac{\rho_s}{2 \rho_0} + i s \right)
  \, .
\end{equation}
%%%%%%%
Assuming further that $u_\nu$ and $v_\nu$ are stationary waves with frequency
$\omega$, we obtain from our solution (\ref{eq:stat}) of the hydrodynamic
sound-wave equation and from the linearized Bernoulli equation (\ref{eq:bern})
the Bogoliubov modes
%%%%%%%
\begin{eqnarray}
  \label{eq:uv}
  u_\nu &\sim& A_\nu \left( \frac{\omega}{2 \alpha z} + \frac{m c }{ \hbar}
  \right) z^{i \omega / \alpha} e^{-i \omega t} \, , \nonumber \\
  v_\nu &\sim& A_\nu \left( \frac{\omega}{2\alpha z} - \frac{m c}{\hbar} \right)
  z^{i \omega/\alpha} e^{-i \omega t} \, .
\end{eqnarray}
%%%%%%%
These asymptotic expressions are valid as long as the elementary excitations
are sound waves with wave number (\ref{eq:kz}).
We show in the Appendix that this is the case sufficiently far away from the
turning points and on either the upper or the lower half plane. Here the term
$\omega/(2 \alpha z)$ in the expressions (\ref{eq:uv}) is always small
compared with $m c / \hbar$. For real and positive $z$ we have
%%%%%%%
\begin{equation}
  \label{eq:pm}
  (-z)^{i\omega/\alpha} = e^{- (\pm 2\pi \omega/\alpha)} z^{i\omega/\alpha}\, .
\end{equation}
%%%%%%%
The $\pm$ sign refers to the two ways in which we may circumvent the
trans-acoustic region, on the upper $(+)$ or on the lower $(-)$ half plane.
Modes with the acoustic asymptotics (\ref{eq:uv}) throughout the
upper half plane are suppressed on the left side of the horizon and for
positive frequencies $\omega$ and enhanced for negative $\omega$.
Modes with the asymptotics (\ref{eq:uv}) on the lower half plane show
the opposite behavior. 

\section{Negative energy}

Bogoliubov modes are normalized according to the scalar products
(\ref{eq:norm}) and (\ref{eq:zero}). Let us calculate the norm of the modes
(\ref{eq:uv}) of the sonic hole. 
The energy parameter (\ref{eq:eps}) is small and so is the
extension of the trans-acoustic region around the horizon, measured roughly
by the location of the turning points (\ref{eq:z0}).
Consequently, we can neglect the trans-acoustic contribution to the
normalization integral (\ref{eq:norm}). 
We approximate the Bogoliubov modes by their asymptotic expressions 
(\ref{eq:uv}), utilize the relation (\ref{eq:pm}), and get
%%%%%%%
\begin{eqnarray}
  \label{eq:integral}
   \int \left( u_\nu^* u_{\nu'} - v_\nu^* v_{\nu'} \right) \, dz 
  &\sim& |A_\nu|^2 \left( \int_{-\infty}^0 + \int_0^{+\infty} \right)
  \frac{2 \omega k_c}{\alpha z} \, z^{i(\omega'-\omega)/\alpha} \, d z
  \nonumber \\
  &= &|A_\nu|^2 \left( 1- e^{- (\pm 2\pi \omega/\alpha)} \right) 4
  \pi \omega k_c\, \delta(\omega'-\omega) \, .
\end{eqnarray}
%%%%%%%
If we choose modes with the asymptotics (\ref{eq:uv}) on the upper half plane the
norm is positive and the $u_\nu$ and $v_\nu$ may serve as proper Bogoliubov
modes. We find the normalized amplitude 
%%%%%%%
\begin{equation}
  \label{eq:a}
  A_\nu = \left[ \left( 1- e^{- 2\pi \omega/\alpha} \right) 4
  \pi \omega k_c\right] ^{-1/2}\, .
\end{equation}
%%%%%%%
Remarkably, the Bogoliubov norm is positive also for modes with negative
frequencies. The Bogoliubov-deGennes equations have a well-known symmetry
\cite{Fetter1,Fetter2}: If the
$(u_\nu,v_\nu)$ are solutions then the complex-conjugated and interchanged
modes, $(v_\nu^*,u_\nu^*)$, are solutions as well. Yet the norm of the
conjugate modes is the negative norm of the original $(u_\nu,v_\nu)$.
In contrast,
sonic black holes generate negative-frequency modes with positive norm, which
is the unusual feature that gives rise to the acoustic analog of Hawking radiation 
\cite{Brout}. We also see how the mentioned symmetry of the 
Bogoliubov-deGennes equations \cite{Fetter1,Fetter2} appears in our case. 
If we chose the $u_\nu$ and $v_\nu$ with
the asymptotics (\ref{eq:uv}) on the lower half plane we would get negative
normalization integrals  (\ref{eq:integral}) for both positive and negative
frequencies.

The negative frequencies of the positive-norm Bogoliubov modes give rise to
negative energies in the Hamiltonian of the elementary excitations,
%%%%%%%
\begin{equation}
  \label{eq:hamiltonian}
  \hat H = \int \hbar \omega \left(\hat a^\dagger_{+} \hat a_{+}
   - \hat a_{-}^\dagger \hat a _{-} \right) N \, d\omega \, .
\end{equation}
%%%%%%%
Here and later the subscripts $\pm$ refer to positive and negative frequencies, 
respectively, and $N$ denotes the density of modes. 
We see that there is no natural ground state of the elementary excitations.
In practice, of course, the spectrum is limited by the requirement (\ref{eq:eps})
that the energies of the excitations ought to be much smaller in magnitude 
than the condensate's mean-field energy.
We note that the Hamiltonian (\ref{eq:hamiltonian}) is invariant
under the Bogoliubov transformations
%%%%%%%
\begin{eqnarray}
  \label{eq:transf}
  \hat a'_{\pm} &=& 
  \hat a_{\pm}\cosh \xi - \hat a_{\mp}^\dagger \sinh \xi
  \, , \nonumber \\
  u'_{\pm} &=& u_{\pm} \cosh \xi - v_{\mp}^* \sinh \xi 
  \, , \nonumber \\
  v'_{\pm} &=& v_{\pm} \cosh \xi - u_{\mp}^* \sinh \xi 
  \, .
\end{eqnarray}
%%%%%%% 
with an arbitrary real parameter $\xi$.
In the case we choose
%%%%%%%
\begin{equation}
  \tanh \xi = e^{-\pi \omega/\alpha}
\end{equation}
%%%%%%%
we get a new set of modes, Eq.\ (\ref{eq:uv}), with
%%%%%%%
\begin{equation}
  \label{eq:prime}
  A'_{\pm} \sim \Theta(\pm z) (4 \pi \omega k_c)^{-1/2} \, .
\end{equation}
%%%%%%%
The $\Theta$ function indicates that the primed modes appear on either
the left or on the right side of the trans-acoustic region. Therefore, 
despite the trans-Planckian problem, a sonic horizon exists,
but at a less well-defined location, within $|z| \lesssim |z_0|$, 
and the horizon applies to a particular set of modes only.

\newpage

\section{Hawking effect}

The Bogoliubov transformations (\ref{eq:transf}) relate
one set of quasiparticles to
another one, both representing perfectly valid energy eigenvalues, yet
their quasiparticle vacua differ (the states $|0 \rangle$ or
$|0'\rangle$ that are annihilated by $\hat a_\pm$ and $\hat a'_\pm$,
respectively). This ambiguity has direct physical consequences,
because the cloud of non-condensed atoms  \cite{LL9}
depends on the vacuum state of the elementary excitations,
%%%%%%%
\begin{eqnarray}
  \langle 0|\, \hat \phi^\dagger \hat \phi \,| 0 \rangle &=& \int 
  \Big( | v_{+} |^2 + | v_{-} |^2 \Big) N \, d \omega\,,
  \label{eq:depletion}\\
   \langle 0'|\, \hat \phi^\dagger \hat \phi \,| 0' \rangle &=& \int 
  \Big( | v'_{+} |^2 + | v'_{-} |^2 \Big) N \, d \omega
  \nonumber\\
  &\neq& \langle 0|\, \hat \phi^\dagger \hat \phi \,| 0 \rangle\,.
\end{eqnarray}
%%%%%%%
In general relativity the notion of the vacuum 
is observer-dependent. For example, the vacuum of empty space in
Minkowski coordinates appears as a thermal field to accelerated
observers \cite{BD,Brout,Unruheffect}. 
In the case of the black hole, the gravitational collapse has created
a state of quantum fields that an inward-falling observer perceives
as vacuum, yet an external observer sees as thermal radiation,
Hawking radiation \cite{BD,Brout,Hawking}.
In our case, the
equivalent of the gravitational collapse, the formation of the sonic
horizon, chooses the quasiparticle vacuum, if the trans-sonic velocity
profile has initially been created from a condensate without a
horizon. Such a process must be sufficiently smooth to keep the
condensate intact.

To analyze the quasiparticle vacuum, we use the Heisenberg picture of
quantum mechanics where observables evolve while the quantum state is
invariant. We describe the initial (and final) vacuum state with
respect to one set of continuous modes given before the formation of
the horizon. In the Heisenberg picture these modes evolve.
We sort the initial modes into left- and right-moving modes that, 
close to complex $\infty$, are analytic on the upper or on the lower half 
plane, respectively, because here $\exp(ikz)$ converges for positive $k$ 
on the upper and for negative $k$ on the lower half plane.
The upstream modes we are interested in stem from right-moving modes.
The formation of the sonic horizon, a smooth process, cannot 
fundamentally alter the analyticity of the vacuum modes. In particular, 
the process can never create non-analytic modes of the type expressed in
Eq.\ (\ref{eq:prime}). Consequently, the initial quasiparticle vacuum 
assumes the analytic modes of Eqs.\ (\ref{eq:uv}) and (\ref{eq:a}).

Given this vacuum state, we determine the quantum depletion of the condensate.
We write the density of the non-condensed atoms in terms of the 
primed Bogoliubov modes. As we have seen, these modes describe the
set of elementary excitations that exhibit the sonic horizon. We find that
%%%%%%%
\begin{equation}
  \langle 0|\, \hat \phi^\dagger \hat \phi \,| 0 \rangle = \rho (|z|) 
  \, , \quad
  \rho(z) = \int 
  \Big[(|u'_{+}|^2 + |v'_{+}|^2)\, \bar{n}(\omega)
     + | v'_{+} |^2 \Big] N \, d \omega\,.
\end{equation}
%%%%%%%
Here $\bar{n}(\omega)$ denotes the average number of non-condensed 
atoms per excitation mode,
%%%%%%%
\begin{equation}
   \bar{n}(\omega)
   = \frac{1}{e^{2\pi\omega/\alpha}-1}
   =\frac{1}{e^{\hbar \omega/k_{_B} T}-1}\,.
\end{equation}
%%%%%%%
The non-condensed atoms are Planck-distributed with the 
Hawking temperature (\ref{eq:hawking}).
Therefore, as soon as the condensate flows through the nozzle,
breaking the speed of sound, a thermal cloud of atoms is formed.
This effect is the signature of Hawking radiation for sonic holes
in Bose-Einstein condensates.
The thermal cloud due to the Hawking effect should be observable 
when the initial temperature of the atoms is below the Hawking 
temperature. 
On the other hand, one could also regard the Hawking effect in 
the condensate as the quantum depletion (\ref{eq:depletion})
of atoms at zero temperature
with respect to the analytic modes (\ref{eq:a}) that transcend the 
horizon. This feature reflects the ambiguity of the vacuum
in general relativity. 
Using techniques for measuring the 
population of Bogoliubov modes \cite{Vogels,Ozeri}, 
one could perhaps demonstrate
the ambiguity of the vacuum in the laboratory.

Finally we note that within our model sonic black holes are stable,
provided of course that the transsonic flow is not plagued by
hydrodynamic instabilities.
In reality, elementary excitations interact with each other,
giving rise to what is known as Landau-Beliaev damping 
\cite{Pitaevskii,Giorgini,Katz}.
Since a sonic horizon does not have a ground state,
this damping mechanism will lead to the gradual evaporation
of the condensate. Therefore, Landau-Beliaev damping 
\cite{Pitaevskii,Giorgini,Katz}
plays the role of black-hole evaporation.
It is tempting to turn matters around and to approach 
cosmological problems from the perspective of 
condensed-matter physics \cite{Chapline,Volovik,Zhang}.

\section*{Acknowledgements}

We thank M. V. Berry, I. A. Brown, L. J. Garay, 
T. A. Jacobson, R. Parentani, and G. E. Volovik
for discussions.
Our work was supported by the ESF Programme 
Cosmology in the Laboratory,
the Leverhulme Trust,
the National Science Foundation of Hungary (contract No.\ F032346),
the Marie Curie Programme of the European Commission,
the Royal Society of Edinburgh,
and by the Engineering and Physical Sciences Research Council.

\newpage

\section*{Appendix}

In this Appendix we examine the asymptotics of the Bogoliubov modes
on the complex plane. We use the semiclassical approximation
\cite{LKO,Csordas}
%%%%%%%
\begin{eqnarray}
   u_\nu &=& {U}_\nu\exp\left(i\int k\,dz  - i\omega t\right)\,,\nonumber\\
   v_\nu &=& {V}_\nu\exp\left(i\int k\,dz  - i\omega t\right)\,.
\end{eqnarray}
%%%%%%%
The wavenumber $k$ should obey Bogoliubov's dispersion relation 
(\ref{eq:bogo}) including the Doppler shift (\ref{eq:doppler}).
We obtain four fundamental solutions of this fourth-order equation.
Figure 2 shows the three branches of $k$ that are relevant in our analysis.
The amplitudes ${U}_\nu$ and ${V}_\nu$ obey the relation
\cite{LKO,Csordas}
%%%%%%%
\begin{equation}
   \label{eq:flux}
   \partial_z\left({U}_\nu^2-{V}_\nu^2\right) v = 0 \,,
\end{equation}
%%%%%%%
where $v$ denotes the group velocity (\ref{eq:group}) 
of the elementary excitation.
Equation (\ref{eq:flux}) formulates the conservation law of the quasiparticle
flux for stationary states \cite{Csordas}
if $z$ is real, where ${U}_\nu^2-{V}_\nu^2$ gives
$|{U}_\nu|^2-|{V}_\nu|^2$ up to a constant phase factor.
The relation (\ref{eq:flux}) can be extended to the complex $z$ plane and to
complex frequencies $\omega$ \cite{LKO}.
Equation (\ref{eq:flux}) implies that the amplitudes ${U}_\nu$ and ${V}_\nu$
diverge close to a turning point $z_0$ where $v$ vanishes.
Consequently, the semiclassical approximation breaks down at the turning point
\cite{LL3}. The turning point causes significant scattering, 
{\it i.e.} the conversion of one mode with a given $k$ into two modes, 
one with wavenumber $k$ and the other one with a different wavenumber 
that satisfies the dispersion relation as well. 
At the turning point the two branches coincide.
To prove this, we regard for a moment $z$ as a function of $k$ at constant 
frequencies $\omega=\omega_0$, defined implicitly by Eqs.\ 
(\ref{eq:doppler}), (\ref{eq:alpha}) and (\ref{eq:bogo}). We get
%%%%%%%
\begin{equation}
v=\frac{\partial \omega}{\partial k}
= u + \frac{\partial}{\partial k}\left(\omega_0 - uk\right)
= -\alpha k \frac{\partial z}{\partial k} \,\,.
\end{equation}
%%%%%%%
We see that the function $z(k)$ reaches extrema at the point $z_0$ 
where $v$ vanishes, {\it i.e.} at the turning point.
Close to the $z_0$ we find after some algebra \cite{LKO}, 
by expanding $z$ into a power series in $\varepsilon^{1/3}$,
%%%%%%%
\begin{equation}
\label{eq:k0}
z-z_0 \sim -\frac{3(k-k_0)^2}{8\alpha}
\,\,,\quad
k_0=\sqrt[3]{-4\varepsilon}
\,\,,
\end{equation}
%%%%%%%
to leading order. 
Consequently, each turning point connects two branches of the
wavenumber $k$. In general the mode conversion occurs near specific lines
in the complex plane, called Stokes lines in the mathematical
literature \cite{Ablowitz}.
Stokes lines, originating from the turning point $z_0$, 
are defined as the lines where the differences 
between the phases $\int k\,dz$ of the two connected $k$ branches 
is purely imaginary. Here one of the waves is exponentially small
compared with the other. We obtain from Eq.\ (\ref{eq:k0})
that the difference between the two branches is proportional to the
square root of $z-z_0$. Consequently, the phase difference is
proportional to $(z-z_0)^{3/2}$, giving rise to three Stokes lines
from each turning point $z_0$, as in the traditional case of 
Schr\"odinger waves in one dimension \cite{LL3,Furry}.
Figure 3 shows the Stokes lines for the three turning points
close to the horizon and for the branch cuts of $k$ chosen in Fig.~2.

At a Stokes line the phase difference between
the two connected branches is purely imaginary.
One of the waves exponentially exceeds the other and,
within the semiclassical approximation,
the smaller wave is totally overshadowed by the larger one,
if the larger wave is present. 
In general, the Bogoliubov modes consist of a superposition of the four 
fundamental solutions that correspond to the four branches of the 
dispersion relation
%%%%%%%
\begin{eqnarray}
  u_\nu &=& c_{_A}u_{_A} + c_{_B}u_{_B} + c_{_C}u_{_C} + c_{_D}u_{_D}
  \,,\nonumber\\
  v_\nu &=& c_{_A}v_{_A} + c_{_B}v_{_B} + c_{_C}v_{_C} + c_{_D}v_{_D}
  \,.
\end{eqnarray}
%%%%%%%
The $u_{_A}$ and $v_{_A}$ refer to the $k$ branch where $k$
obeys the asymptotics (\ref{eq:kz}), {\it i.e.} where the wavenumber
satisfies the dispersion relation  (\ref{eq:disrel}) of sounds 
in moving fluids, taking into account the Doppler detuning (\ref{eq:doppler})
and where $k$ corresponds to an upstream wave.
We call such Bogoliubov modes acoustic modes.
When crossing a Stokes line, the exponentially suppressed solution 
may gain an additional component that is proportional to the coefficient
of the exponentially enhanced solution \cite{Furry}.
If we wish to construct Bogoliubov modes where only the exponentially
smaller component exits in the vicinity of a Stokes line 
we must put the coefficient of the larger one to zero. 
In Fig.\ 3 the pairs of letters indicate which branches are connected by the lines,
and the first letter identifies the exponentially dominant branch. 
The picture shows that with the choice of branch cuts made we can 
construct a Bogoliubov mode that is acoustic on the upper half plane.
Trans-acoustic physics is confined to the lower half plane.
On the other hand, if we chose other branch cuts of $k$ we may 
get Bogoliubov modes that are acoustic on the lower half plane
and trans-acoustic on the upper one.
Therefore, according to Eq.\ (\ref{eq:pm}),  
the choice of the $k$ branch determines whether a Bogoliubov mode 
is larger or smaller beyond the sonic horizon at the real axis, for $z<0$.
Branch cuts of $k$ are fairly arbitrary.
Given the Bogoliubov mode on the right side of the horizon,
we cannot predict within the semiclassical approximation the 
amplitude of the mode on the left side.
Therefore, the two sides are causally disconnected. 
Within the semiclassical approximation, the horizon is a genuine
horizon, despite the acoustic analog of the notorious 
trans-Planckian problem \cite{T,J}.

%%%
\begin{figure}
\begin{center}
\includegraphics[width=10cm]{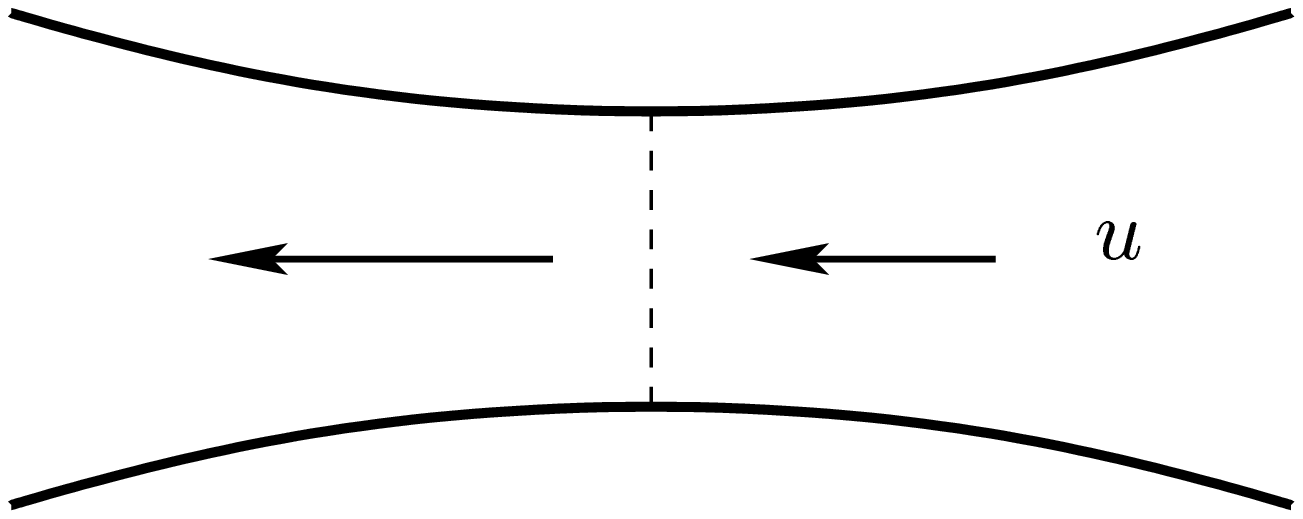}
\end{center}
\caption{\label{fig:nozzle} Schematic diagram of a sonic horizon. 
A fluid is forced to move through a constriction where the flow speed
$u$ becomes supersonic (dashed line). 
The constriction may be formed by the walls of a tube or, 
if the fluid is an alkali Bose-Einstein condensate, 
by a suitable trapping potential.
The picture shows a Laval nozzle \cite{Courant} where 
a supersonic fluid is hydrodynamically stable.}
\end{figure}
%%%

%%%%%%%
\begin{figure}
\includegraphics{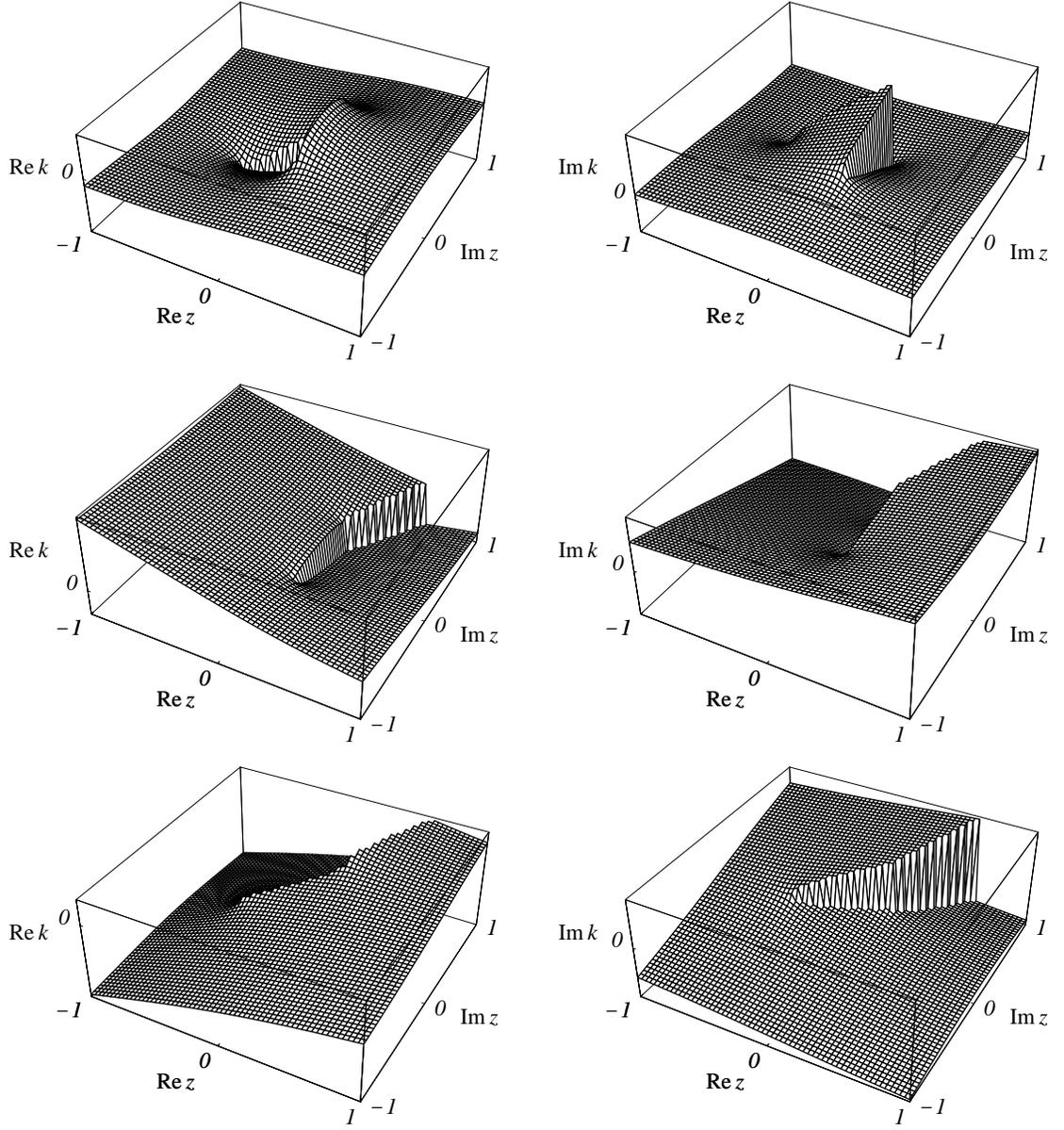}
\caption{\label{fig:branches}
Wavenumbers $k$ of elementary excitations 
around a sonic black-hole horizon, 
analytically continued onto the complex plane.
The figure shows three roots of the dispersion relation
$[\hbar^2k^2/(2m) + mc^2]^2 -
\hbar^2[\omega-(-c+\alpha z)k]^2 =m^2c^4$ for
$\omega=0.1 \,(mc^2/\hbar)$ 
and $\alpha=0.5 \,(mc^2/\hbar)$, 
illustrating the branch cuts of $k$.
The top row displays the wavenumber of a sound wave 
that propagates against the current. 
The picture indicates the characteristic $\omega/(\alpha z)$ 
asymptotics away from the branch points. 
The two lower rows display two 
trans-acoustic branches of $k$. 
The fourth root of the dispersion relation is not shown,
because it corresponds to the trivial case of
sound waves that propagate with the flow.}
\end{figure}
%%%%%%%

%%%%%%%
\begin{figure}
\includegraphics[width=10cm]{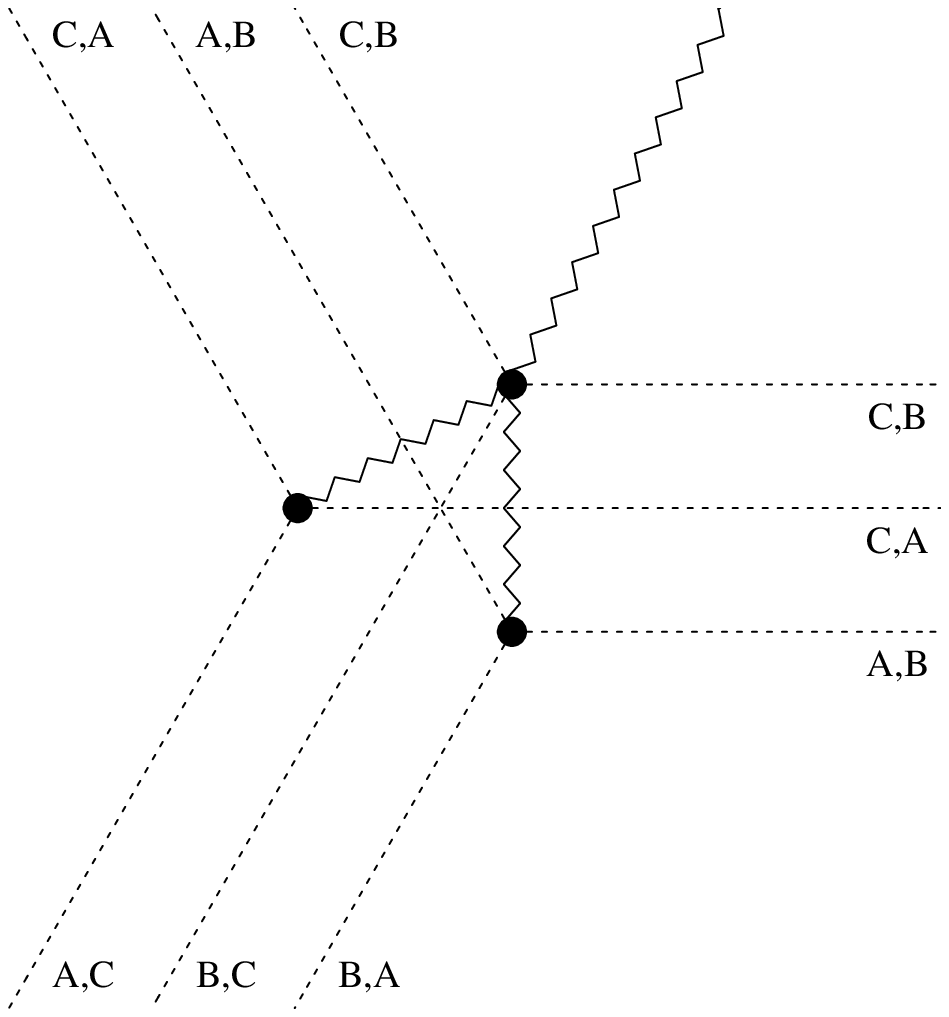}
\caption{\label{fig:stokes1} 
Stokes lines of elementary excitations at a sonic black hole (dotted lines),
given the choice of branch cuts made in Fig.\ 2.
The pairs of letters indicate which branches of the superposition (52)
are connected by the lines. 
The first letter of each pair identifies the exponentially dominant branch.
We construct a Bogoliubov mode that is acoustic (component A) 
on the upper half plane by putting the coefficient $c_{_C}$ to zero on the
C,A Stokes line originating from the left turning point.}
\end{figure}
%%%%%%%


\begin{thebibliography}{99}
\bibitem{Remark2}
In a fluid extending in more than one spatial dimensions
the acoustic space-time geometry in curved, in general.
There is no universal comoving frame anymore,
because in this frame the metric is flat.
\end{thebibliography}
\end{document}